\documentclass[twocolumn]{openjournal}

\usepackage{bm}
\usepackage{bold-extra}
\usepackage{xspace}
\usepackage{xcolor}
\usepackage{amsmath}
\usepackage{graphicx}
\usepackage[hidelinks]{hyperref}
\hypersetup{breaklinks=true,colorlinks=true,linkcolor=blue,urlcolor=blue,citecolor=blue}
\usepackage[caption=false]{subfig}
\usepackage{booktabs}
\usepackage[capitalise]{cleveref}

\usepackage{savesym}
\savesymbol{tablenum}
\usepackage{siunitx}
\restoresymbol{SIX}{tablenum}

\usepackage{tabularx}

\usepackage{lineno}

\usepackage{ragged2e}

\newcommand{\hii}{\relax \ifmmode {\mbox H\,{\scshape ii}}\else H\,{\scshape ii}\fi}
\newcommand{\mi}{\relax \ifmmode {\mu{\mbox m}}\else $\mu$m\fi}
\newcommand{\ha}{\relax \ifmmode {\mbox H}\alpha\else H$\alpha$\fi}
\newcommand{\hb}{\relax \ifmmode {\mbox H}\beta\else H$\beta$\fi}

\newcommand{\sii}{\relax \ifmmode {\mbox S\,{\scshape ii}}\else S\,{\scshape ii}\fi}
\newcommand{\siii}{\relax \ifmmode {\mbox S\,{\scshape iii}}\else S\,{\scshape iii}\fi}
\newcommand{\siv}{\relax \ifmmode {\mbox S\,{\scshape iv}}\else S\,{\scshape iv}\fi}
\newcommand{\nii}{\relax \ifmmode {\mbox N\,{\scshape ii}}\else N\,{\scshape ii}\fi}
\newcommand{\oi}{\relax \ifmmode {\mbox O\,{\scshape i}}\else O\,{\scshape i}\fi}
\newcommand{\oii}{\relax \ifmmode {\mbox O\,{\scshape ii}}\else O\,{\scshape ii}\fi}
\newcommand{\oiii}{\relax \ifmmode {\mbox O\,{\scshape iii}}\else O\,{\scshape iii}\fi}
\newcommand{\neiii}{\relax \ifmmode {\mbox Ne\,{\scshape iii}}\else Ne\,{\scshape iii}\fi}

\newcommand{\rdostres}{\relax \ifmmode {\,\mbox{R}}_{\rm 23}\else \,\mbox{R}$_{\rm 23}$\fi} 

\newcommand{\ciii}{\relax \ifmmode {\mbox O\,{\scshape iii}}\else C\,{\scshape iii}\fi}
\newcommand{\civ}{\relax \ifmmode {\mbox O\,{\scshape iii}}\else C\,{\scshape iv}\fi}
\newcommand{\nv}{\relax \ifmmode {\mbox N\,{\scshape v}}\else N\,{\scshape v}\fi}
\newcommand{\heii}{\relax \ifmmode {\mbox He\,{\scshape ii}}\else He\,{\scshape ii}\fi}

\newcommand{\gsim}{\hbox{\rlap{\lower.55ex\hbox{$\sim$}} \kern-.3em
\raise.4ex \hbox{$>$}}}
\newcommand{\lsim}{\hbox{\rlap{\lower.55ex\hbox{$\sim$}} \kern-.3em
\raise.4ex \hbox{$<$}}}

\shorttitle{Optical sulphur abundances based on models}
\shortauthors{P\' erez-Montero et al}

\begin{document}

\journalinfo{The Open Journal of Astrophysics}

\title{Sulphur abundances in star-forming regions from optical emission lines: A new approach based on photoionization models consistent with the direct method}

\author{E. P\'erez-Montero$^{1,\ast}$, B. P\'erez-D\'{\i}az$^{1,2}$, J.~M. V\'{\i}lchez$^1$, I.~A. Zinchenko$^{3,4,5}$, A. Castrillo$^{6,7}$, M. Gavil\'an$^6$, S. Zamora$^{6,7,8}$, and A.~I. D\'{\i}az$^{6,7}$ }
\email{$^\ast$epm@iaa.es}

\affiliation{
$^1$Instituto de Astrof\'\i sica de Andaluc\'\i a. CSIC. Apartado de correos 3004. 18080, Granada, Spain\\
$^2$ INAF - Osservatorio Astronomico di Roma, Via Frascati 33, Monte Porzio Catone, 00040, Italy\\
$^3$Faculty of Physics, Ludwig-Maximilians-Universit{\"a}t, Scheinerstr.1, 81679 Munich, Germany \\
$^4$Main Astronomical Observatory, National Academy of Sciences of Ukraine, 27 Akad. Zabolotnoho St 03680 Kyiv, Ukraine\\
$^5$ Astronomisches Rechen-Institut, Zentrum f\"{u}r Astronomie der Universit\"{a}t Heidelberg, M\"{o}nchhofstra{\ss}e 12-14, D-69120 Heidelberg, Germany \\
$^6$Departamento de F\'{i}sica Te\'{o}rica, Universidad Aut\'{o}noma de Madrid, Madrid 28049, Spain\\
$^7$CIAFF, Universidad Autónoma de Madrid, 28049 Madrid, Spain\\
$^8$Scuola Normale Superiore, Piazza dei Cavalieri 7, I-56126 Pisa, Italy\\
}

\begin{abstract}
The derivation of sulphur chemical abundances in the gas-phase of star-forming galaxies is explored in this work, using the emission lines produced in these regions in the optical part of the spectrum and by means of photoionization models. We adapted the code {\sc HII-CHI-mistry} to account for these abundances by implementing additional grids of models that assume a variable sulphur-to-oxygen abundance ratio, beyond the commonly assumed solar value. The addition of these models, and their use in a new iteration of the code  allows us to use sulphur lines to precisely estimate the sulphur abundance, even in the absence of auroral lines. This approach aligns with the results from the direct method, and no additional assumptions about the ionization correction factor are needed, as the models directly predict the total sulphur abundance.
We applied this new methodology to a large sample of star-forming regions from the MaNGA survey, and we explored the variation of the S/O ratio as a function of metallicity, making corrections for the significant contribution from diffuse ionized gas, which particularly affects the [\sii] emission.
Our results indicate no significant deviations from the solar S/O value in the range 8.0 $<$ 12+log(O/H) $<$ 8.7, where the bulk of the {\sc MaNGA} sample stays, but also with possible enhancements of sulphur at the high metallicity regime. This may be linked to the depletion of oxygen in the gas-phase due to its incorporation onto dust grains, as it remains when other metallicity indicators independent of this depletion, as S/H itself, are used instead. 
\end{abstract}

\maketitle



\section{Introduction}\label{sec:intro}

Sulphur (S), while not the most abundant element, ranks among the most prevalent metals in the Universe. Sulphur is primarily produced through nucleosynthesis via the alpha-chain in massive stars that later eject it into the interstellar medium (ISM). Consequently, S/H serves as a tracer for the total metal content in the gas-phase of galaxies and, when compared with other species, such as oxygen (O), it also provides insights into the processes of production, ejection, and mixing within the gas-phase.

Given its main channel of production and mass range of the stars creating it, S abundance is usually assumed  to scale with O abundance following solar proportions (i.e. log(S/O)$_{\odot}$ = -1.57, \citealt{asplund09}), which is consistent with the results  obtained in  several works (e.g. \citealt{garnett97,izotov06,chaos-n3184}). 

However, despite this general consistency, other research supports a constant S/O ratio slightly below the solar value \citep{kehrig06}, and even some studies have found non-negligible deviations from this ratio at distinct metallicity regimes.
For instance, some authors find evidence of an increasing S/O ratio with decreasing metallicity (e.g. \citealt{vilchez88, dors16,zinchenko24}),
what could be explained by e.g., including pair-instability supernovae (PISN) in chemical modeling \citep{goswami24}.
Moreover, the observed behavior of S/O as a function of metallicity could be even bimodal, according to \cite{dz22}, as,
besides the  high S values found in the metal-poor regime and after finding low S/O at solar metallicity,  also report an increase of S/O for higher values, this latter possibly related with the different refractory nature of O and S \citep{henry93}.
This result, with very low S/O values in some galaxies with solar-like oxygen abundances has also been reported by \cite{pd24} based on sulphur infrared emission lines.
The determination of S abundances in the gas-phase of galaxies in the optical and near-IR ranges, the unique so far for which a statistically significant amount of objects 
can be studied to explore S/O at all regimes,
can be made by means of several bright emission lines of collisional nature (CELs) (i.e. [\sii] $\lambda$ 6717,31 \AA\, [\siii] $\lambda$ 9069,9532 \AA). On one hand, the relative abundances of the 
ions responsible for the emission of these lines can be accurately determined following the direct or $t_e$ method \citep{pm17}, if, in addition, their respective auroral lines (i.e. [\sii] 4068,4072 \AA, [\siii] 6312 \AA) can also be measured with enough confidence to allow for a determination of the corresponding electron temperatures. 

However, one of the main drawbacks of the optical  regime to be used for the derivation of S abundance stays on the fact that a non-negligible fraction of the sulphur abundance can be found in higher stages of ionization,
such as S$^{+3}$, only measurable in the infra-red regime (e.g. at [\siv] $\lambda$ 10.5 $\mu$m) \citep{vermeij02}. The acquisition of mid-IR data could thus be undertaken to have a more precise quantification of the total sulphur abundance, above all in nebula ionized by very hard incident radiation fields, such as Extreme Emission Line Galaxies (EELGs) or Active Galactic Nuclei (AGN).
However, if the observed spectrum is not complete,  even if the direct method can be applied, additional assumptions must be taken to calculate the total S abundance by means of an ionization correction factor (ICF) to account for the sulphur stages unseen in the optical or IR parts, contributing to enhance the uncertainty of the S abundance derivation \citep{stasinska78,pm06,dors16}.

In addition, despite the fact that the emissivity of the auroral line from [\siii] at $\lambda$ 6312 \AA\, depends less on temperature
than other optical auroral lines, such as [\oiii] $\lambda$ 4363 \AA, 
, what makes easier to detect it at higher metallicity (e.g. \citealt{castellanos02,bresolin05}),
the direct method for the determination of S abundances can hardly be applied in large samples composed of weak/distant objects. Therefore, the estimation of S chemical abundance in the gas-phase must rely on the application of other techniques based on strong detected emission-lines.
For instance, one of the most widely used parameters based uniquely on strong sulphur CELs is the $S23$ parameter, defined by \cite{ve96} as:

\begin{equation}
S23 = \frac{\mathrm{[S\,II]} 6717,6731 \AA + \mathrm{[S\,III]} 9069,9532 \AA}{H\beta}
\end{equation}

This parameter was calibrated to provide an estimate of the total oxygen abundance (\citealt{dpm00,pmd05}), and also as a direct tracer of the total sulphur abundance by \cite{pm06}. However, the fact that the same parameter can be used either as a tracer for O, or for S, underlines the basic assumption of a constant S/O, so it cannot be used to figure out the variation of S/O with metallicity.

Another alternative to calculate abundances is the use of photoionization models, that when tailored to individual objects can account for the calculation of either ICF when auroral lines are detected, or the total S abundance from the unique use of strong lines, (e.g. \citealt{evans86,garnett89,pm07,pm10}), but only for a limited number of objects.
On the other hand, the program {\sc HII-CHI-mistry} (hereinafter {\sc HCm} \citealt{hcm14})
uses another strategy to provide an estimate of the total elemental abundances based on a limited set of emission-lines with enough signal-to-noise ratio that can be applied to large number of objects.

There is a version of this code for the IR ({\sc HCm-IR}, \citealt{jafo21,pd22,pd24}), that is already prepared to calculate S abundances, taking advantage from the fact that S lines in the rest frame mid-IR have very low dependence on electron temperature, what enables the code to independently estimate O and S abundances. However, the situation in the optical regime is quite different, as S CELs are much more dependent on electron temperature.
For this reason, in this paper we describe a new version of {\sc HCm} for the optical rest frame spectral range to provide solutions for the S chemical abundance consistently with the direct method that can be applied even in absence of any auroral line, and we applied it to a large sample of star-forming regions in the MaNGA survey to shed some light on the behaviour of the S/O ratio with metallicity. 

The paper is organized as follows: In Section 2, we describe the observational data compiled from several sources, including one sample with direct determinations of the sulphur abundance, used to evaluate the results of our code, and another larger sample to which the code was applied. In Section 3, we explain the method, detailing the grids of photoionization models and the {\sc HCm} code. In Section 4, we present and discuss our results. Finally, in Section 5, we summarize our findings and present our conclusions.

\section{Data sample}\label{sec:data}

For this work we compiled two main sources of observational information. The first sample has the aim of serving as a control sample for our methodology gathering optical spectral data, allowing a direct determination of their total sulphur abundance that can be used  to evaluate the results of our code.
The second one is a sample of pointings from the {\sc MaNGA} survey selected as ionized mainly by star-formation. Although the quality of this {\sc MaNGA} data is good, the observations are not deep enough to derive total abundances based on the application of the direct method, relying on the detection of the weak auroral lines. Therefore, the application of other methods exclusively based on the use of strong emission lines is necessary.  

\subsection{Control sample of objects with direct sulphur abundance}

The application of the direct method to derive total oxygen and sulphur abundances based on relative fluxes of CELs in the optical range relies on the detection of any auroral line whose ratio with other stronger nebular line emitted by the same ion permits the derivation of the electron temperature.
In addition, the determination of S abundances requires the detection of the lines emitted by the most abundant ions in the gas (e.g. $S^+$ and $S^{2+
}$) and the application of an ICF to account for the fraction of S in unseen higher ionization stages, such as $S^{3+}$. 

Therefore, we used the data compiled by \cite{dz22}, who gather reddening-corrected emission-lines fluxes of optical lines of [\oii], [\oiii], [\sii], and [\siii], with additional auroral lines for [\oiii] and [\siii], for the calculation of total oxygen and sulphur chemical abundances. This sample includes 255 disk \hii\ regions (DHR) and 97 \hii\ galaxies (HIIG), extending over a wide range of metallicity.

We also took for our analysis the total S abundances calculated in this work, using ICFs based on the softness parameter \citep{vp88} or, in the case of observations only in the red part of the spectrum, based on Ar emission-lines. However, for O abundances, given that these are not provided in the same work for the whole sample, we recalculated them from the listed emission-lines, using {\sc pyneb} \citep{pyneb}, and adopting the empirical relation between oxygen and sulphur electron temperatures given by {\cite{hagele06}, following the procedure for the direct method described in \cite{pm17}.

\subsection{The MaNGA sample}

The second sample of data was compiled in order to largely increase the database of objects with a determination of the total S abundance with no need of a direct determination based on the electron temperature, but with optical strong [\sii] and [\siii] emission lines.

A sample that fulfills these conditions  corresponds to the different spaxels representing star-forming regions selected from  the Mapping Nearby Galaxies at Apache Point Observatory \citep[MaNGA;][]{Bundy2015}, which is part of the Sloan Digital Sky Survey IV \citep[SDSS IV;][]{Blanton2017}.
The use of {\sc MaNGA} allows us to  study a sample statistically significative to analyze the behaviour of the fundamental relations between abundances in a wide range of variation without the constrain of having auroral lines, what enourmously expand the number of objects for which the derivation of abundances can be applied and, can hence result an ideal scenario to make use of  our methodology.

As in \cite{pm23},  we took the data from  the {\sc MaNGA} data release 17 (DR17, \citealt{sdss-dr17}), from which we selected 2124 galaxies in the redshift range 0.0005 $< z <$ 0.0845 where both [\siii] $\lambda\lambda$9069,9532 \AA\ lines are available in the MaNGA wavelength range.
We selected those  spaxels in the corresponding datacubes with a size significantly smaller than the point spread function (PSF). As the spatial resolution of these
cubes has a median  full width at half maximum (FWHM) of 2.54 arcsec \citep{Law2016}, this corresponds to an average 
spatial resolution of 1.2 kpc, considering a  median $z$ of 0.024.

For each selected spaxel, we collected emission-line fluxes  with at least  a signal-to-noise ratio (S/N) of 10 for [\oii] $\lambda$3727 \AA, [\oiii] $\lambda\lambda$4959,5007 \AA, [\nii] $\lambda$6584 \AA, [\sii] $\lambda\lambda$6717,6731 \AA, and [\siii] $\lambda\lambda$9069,9532 \AA. Such high S/N has been chosen to minimize false detections of [\siii] $\lambda\lambda$9069,9532 \AA\ lines due to posible telluric contamination of the spectrum. Moreover, to minimize chance of false detection, we allowed fitting of  [\siii] $\lambda\lambda$9069,9532 with independent velocity dispersion of each line in doublet and then selected spectra with differences in the velocity dispersion less then 30 km/s. This implies a total of 201 735 spectra. To obtain the emission line fluxes, we used the {\sc STARLIGHT} code \citep{CidFernandes2005,Mateus2006,Asari2007} to subtract the stellar background and the {\sc ELF3D} code to fit the emission lines. Details about the processing can be found in \citet{zinchenko2016,z21}.  
The line fluxes were corrected for interstellar reddening using the analytical
approximation of the Whitford interstellar reddening law \citep{Izotov1994},
assuming the Balmer line ratio of $\text{H}\alpha/\text{H}\beta = 2.86$.
When the measured value of $\text{H}\alpha/\text{H}\beta$ is less than 2.86,
the reddening is set to zero.

\section{The models and the code}\label{sec:code}

In this section we describe the code {\sc HCm} to calculate chemical abundances and the ionization parameter, $U$, from a set of observed emission lines as compared to the predictions from a large grid of photoionization models. We focus on the new version of the code, version 6.0, capable of providing an estimate of the total sulphur abundance using optical [\sii] and [\siii] lines.  We also describe the new considered observables based on these lines, the new added grid of models with a variable S/O ratio,  and the calculation procedure.

\subsection{Description of the photoionization models}

In order to derive abundances, {\sc HCm} operates using different grids of photoionization models defined by the user, but it incorporates by default libraries of emission line fluxes calculated with the code  {\sc Cloudy} v. 17 \citep{cloudy}.
As in previous versions of the code (e.g. \citealt{hcm14, pm21}), for the derivation of O/H, N/O, and $U$ the models calculated for star-forming regions consider input conditions covering a range for 12+log(O/H) from 6.9 to 9.1 in bins of 0.1 dex, for log(N/O) from -2.0 to 0.0 in bins of 0.125 dex, and for log $U$ from -4.0 to -1.5 in bins of 0.25 dex.
These models assume as incident radiation source {\sc Popstar} synthetic model atmospheres from \cite{popstar} with an Initial Mass Function (IMF) from \cite{chabrier} with an age of 1 Myr. The gas distribution is assumed to have  plane-parallel geometry, constant electron density of 100 cm$^{-3}$, a filling factor of 0.1, and a standard galactic dust-to-gas mass ratio.

In addition, as a novelty for this version and departing from the photoionization model grids already calculated for previous versions of the code, we also  used {\sc Cloudy} v. 17 to complement these grids varying in each model different input values of S/O, considering deviations from the solar ratio previously assumed in the original grid, but with the same values for the rest of input conditions.
Then,  for each combination of O/H and $U$, we calculated  six additional models with different S/O ratio ranging from a deviation of -0.6 to 0.6 dex in relation to the solar value (log(S/O)$_{\odot}$ = -1.57, \citealt{asplund09}) in bins of 0.2 dex. The rest of elements were still fixed to solar proportions according to the values given by \cite{asplund09}, with the exception of N, which was fixed in these additional models according to the empirical relation with O/H found by \cite{hcm14}. Note that this is not incompatible with the N/O variation considered in previous models, as this does not imply noticeable changes in the derivation of S/H. Overall, this additional grid with variable S/O implies a number of 1\,518 models.

These additional models with varied S/O were also incorporated to other libraries of model grids available for  the code, including BPASS v.2.1 \citep{bpass} models for its use for EELGs as described in \cite{pm21}.
However, the inclusion of these new models and their use for the calculation of S/O has not been made for its use in  the Narrow Line Regions (NLR) in AGN \citep{hcm-agn,pd25}
but it will be incorporated in the near future and discussed in more detail in forthcoming papers.  
Thus, for the above described types of ionizing source, mainly representing star-forming objects, the code incorporates two libraries listing the predicted emission line fluxes and the introduced initial conditions and abundances: one with solar S/O ratio for varying O/H, N/O, and $U$; and another one assuming an empirical relation between O/H and N/O with varying S/O for each value of  O/H, and $U$. 

\begin{figure*}
   \centering
\includegraphics[width=0.4\textwidth,clip=]{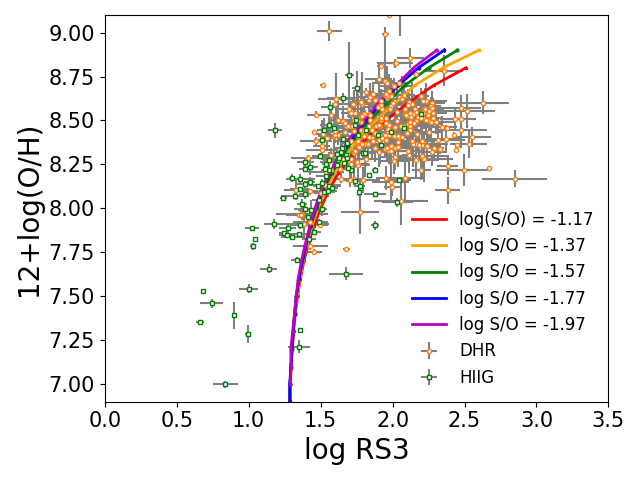}
\includegraphics[width=0.4\textwidth,clip=]{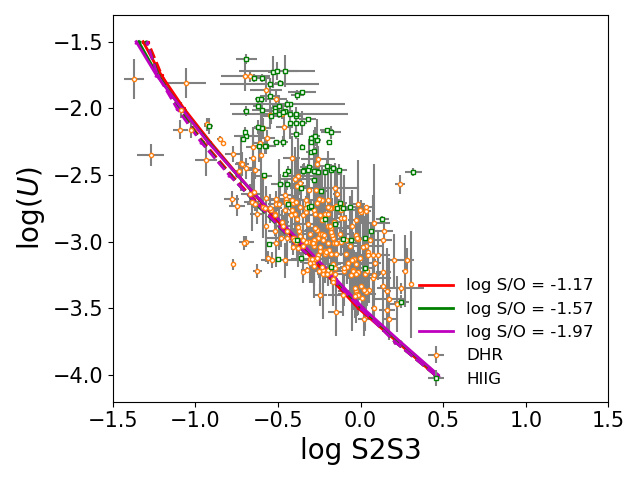}

\caption{Left: Relation between the emission-line ratio $RS3$ with the total oxygen abundance both for the control sample and for different model sequences calculated at a fixed log $U$ = -2.5 and different values of S/O. Right: Relation between the emission line ratio $S2S3$ and log $U$ both for the control sample, as calculated using {\sc HCm}, and for different model sequences calculated at a fixed 12+log(O/H) = 8.7 (8.1) with solid (dashed) line and different values for S/O.}
\label{RS3}%
\end{figure*}

\begin{figure*}
   \centering

\includegraphics[width=0.4\textwidth,clip=]{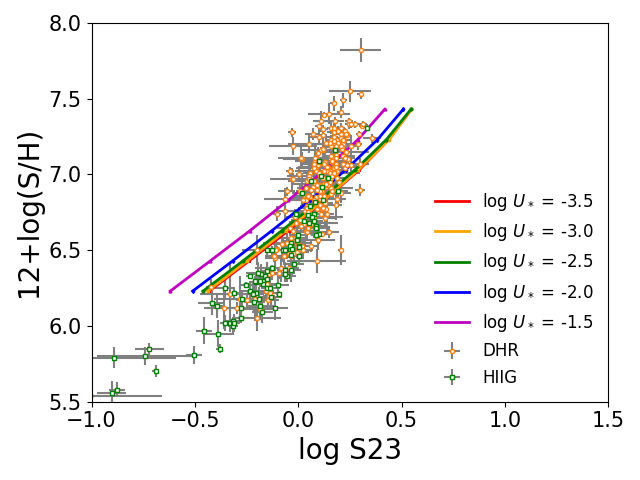}
\includegraphics[width=0.4\textwidth,clip=]{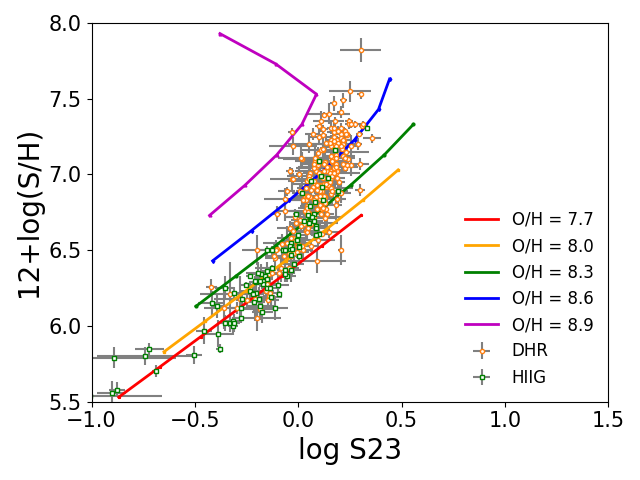}

\caption{Relation between the $S23$ parameter and the total sulphur abundance both for the control sample and for different model sequences. Left: Models calculated at a fixed 12+log(O/H) = 8.4 and varying log $U$. Right: Fixed log $U$ = -2.5 and different values for the total oxygen abundance.}
\label{S23}%
\end{figure*}

\subsection{Description of the code {\sc HII-CHI-mistry} to derive sulphur abundances}

{\sc HCm} \citep{hcm14} is a python-based script aimed at the derivation of certain chemical abundances and the ionization parameter $U$ in ionized gaseous nebulae,  including star-forming regions, AGN, \citep{hcm-agn} or EELGs \citep{pm21}. There are also specific versions able to deal with ultraviolet lines \citep{pma17,pm23} and the IR \citep{jafo21,pd22}\footnote{All the different version of the program can be retrieved from the webpage \url{http://home.iaa.es/~epm/HII-CHI-mistry.html}}.
This calculation is made from the comparison of the measured reddening-corrected fluxes of certain CELs  with the prediction of large grids of photoionization models. The code basically estimates these quantities as the mean values of a $\chi^2$-weighted distribution of all considered models, where the $\chi^2$ are the normalized quadratic difference between some observed and measured emission-line ratios sensitive to the searched quantities.
The fact that our method finds these solutions as compared with    the total abundances taken as input in each model of the used grids is an advantage in relation to the direct method,
as it implies that the calculation of ICFs is not necessary, step that it is often based on arbitrary assumptions made on biased samples or photoionization models that are not necessaryly consistent with the rest of calculations. 
This advantage also extends to the fact that the code, contrary to the direct method, does not need to assume any arbitrary relation between
the electron temperatures associated to the derivation of the corresponding chemical abundance of each ion, as the thermal structure  is implicit in each model, and the ionic abundances are not necessary to calculate the elemental abundances in our method.

The version 6.0 of {\sc HCm} for the optical makes use of the model libraries described above, compiling the   predicted emission-line fluxes and the model parameters required to calculate 12+log(O/H), log(N/O), log $U$, and 12+log(S/H) with their corresponding uncertainties,
estimated considering two independent sources of error: the standard deviation of the final $\chi^2$-weighted distribution, and the uncertainty associated to the given observational errors, calculated performing a Monte Carlo iteration perturbing the fiducial given input values with the uncertainties.

The set of emission lines accepted by the code in this version includes \oii] 3727 \AA\ (sum of the doublet 3726+3729), [\neiii] 3868 \AA, [\oiii] 4363 \AA, [\oiii] 4959,5007 \AA\ (due to the theoretical ratio between these lines, the code admits only one of them), [\nii] 5755 \AA, [\siii] 6312 \AA, [\nii] 6584 \AA, [\sii] 6716,6731 \AA\ (provided separately or summed, but total flux must be specified), [\oii] 7325 \AA\ (total flux of multiplet 7319-7330), and one or both of the doublet lines [\siii] 9069, 9532 \AA. These fluxes must be reddening-corrected and given relative to \hb.

The procedure and decission tree of the code has been designed as follows: The code firstly performs an iteration through the entire grid of models to calculate N/O, as the used observables to derive this ratio do not depend on $U$.
   For this version, the parameter N2O2 (defined as log([\nii]/[\oii]), is used as tracer of N/O by \cite{pmc09} with priority, but N2S2 (similar replacing [\oii] with [\sii])\citep{pmc09} and O3N2 \citep{florido22} can be also used.

Once N/O is fixed in the grid of models, the code performs a second iteration to provide an estimation for 12+log(O/H) and log $U$. In this case, several observables can be used, depending on which of them are available as a function of the provided emission lines for each object. These are all described in \cite{hcm14}. Notice that, once N/O has been fixed in the previous step, the code constrains the grid to the found N/O values, and the observables based on [\nii] can then be used without any previous arbitrary assumption on the O/H vs N/O relation. Nevertheless, in case that a previous N/O estimation has not been possible, an empirical relation between O/H and N/O is adopted instead, although this may imply additional sources of uncertainty if the studied object does not follow it. In addition, the code also constrains the grid in case that any auroral to nebular emission line ratio (e.g. [\oiii] 4363/5007 \AA) is not given, following an empirical relation between O/H and $U$, that will be discussed in next section.

For this version of the code dealing with S emission lines, in this step, in addition to the observables based on the available optical CELs to build the resulting $\chi^2$-weighted distributions to derive O/H and $U$, we added those based on [\siii] emission lines that were not previously used in other versions of the code. Even though these lines present a certain dependence on the relative S abundance, some among them  have a larger dependence on the total metallicity or on excitation.

This is the case of the parameter $RS3$, which is basically the ratio of the fluxes of the nebular-to-auroral [\siii] emission lines:

\begin{equation}
RS3 = \frac{\mathrm{[S\,III]} 9069,9532 \AA}{\mathrm{[S\,III]} 6312 \AA}
\end{equation}

and that is is usually employed as an estimator of the electron temperature in the intermediate ionization region, where ions such as $S^{2+}$ or $Ar^{2+}$ dominate.
Therefore, as the value of the electron temperature is mainly driven by the total amount of metals in the gas-phase, as the gas cooling is mainly produced via the CELs of the ions, its value mostly depend on  total metallicity, more than in a specific ionic or elemental abundance. In left panel of Figure \ref{RS3} we show the behaviour of this parameter as a function of total oxygen abundance, being this a proxy of the total metallicity of the gas, as O is the most abundant element in this phase. The variations of the S/O ratio do not imply large differences on this behaviour, As sulphur is in all cases much less abundant than oxygen, it does not significantly affect gas cooling. Therefore, when this ratio can be used as an input by the code, it is accounted for the calculation of O/H rather than for   the calculation of the total sulphur abundance. This is also the case for other emission-line ratios of nebular to auroral lines not originally included in the code, but that are now considered, such as for [\nii] or [\oii], helping to provide a more precise determination of the total oxygen abundance when these ratios can be measured with enough confidence. However, it is also important to point out that a large associated uncertainty of these ratios used in the input can have the contrary effect if they increment the noise of the final solutions.  

Another widely used parameter based on [\siii] lines that it is usually used for the derivation of physical properties or abundances in ionized gaseous nebulae, is the $S2S3$ parameter, defined as:   

\begin{equation}
S2S3  =  \frac{\mathrm{[S\,II]} 6717,6731 \AA}{\mathrm{[S\,III]} 9069,9532 \AA}
\end{equation}

which has been mainly used as a proxy for the ionization parameter of the gas \citep{diaz98}. In right panel of Figure \ref{RS3} we show the behaviour of the parameter as a function of log $U$ for different model sequences of the grid calculated for two values of 12+log(O/H) = 8.1 and 8.7, and showing also results for different S/O ratios around the solar value. As it can be seen, little variation is found as a function of the relative sulphur abundance and, consequently, this observable is only used in the second iteration of the code, which is aimed at the simultaneous search of both $U_*$ and O/H. Notice that this ratio has been also studied to derive the equivalent effective temperature of the incident field of radiation by \cite{hcm-teff}, as part of the input for the code {\sc HCm-Teff}, so a more detailed determination of the ionization parameter should be made using it if a more precise grid attending to possible variations of the incident SED should be made. This partially explains the dispersion obtained for the studied objects in this plot, as the O2O3 ratio is also used for the derivation of $U$.

When O/H and $U$ have been estimated, with their corresponding errors, the code performs a third iteration in order to estimate S/H, but only using the new grids with a varied S/O ratio, once constrained to the values previously estimated. In this case, the only parameter used to obtain the $\chi^2$-weighted mean from which the value of S/H is derived is $S23$, as defined above in equation (1), although alternative forms of this parameter are also utilized by the code when the nebular [\siii] lines are not given (e.g. using [\siii] at 6312 \AA). 

The basis of this procedure is well justified by inspecting the behaviour of the model sequences shown in Figure \ref{S23}, where $S23$ is mainly driven by S/H, but with a large dependence on both O/H and $U$. Therefore, an independent determination of these parameters prior to the derivation of S/H is necessary to supply a trustable and robust determination of the S/H ratio using the available S emission lines in the optical part of the spectrum.

As a first sanity check of the results obtained by the code, we performed a test consisting of using as input the emission-line fluxes predicted by the models to find out if the code obtains the same chemical abundances used by the models.
The results of this operation are shown in Table \ref{test}, for which we give the mean offsets and the standard deviations between the  calculated O/H and S/H and the corresponding input values using the {\sc popstar} models. We explored different scenarios including  calculations made using: all possible lines, both nebular and auroral; using only strong nebular lines; both auroral and nebular but only in the red part of the spectrum (i.e. from [\siii] $\lambda$ 6312\AA); and, finally, only strong lines in this same regime. We provide results when no error is assigned to the input fluxes and, between parenthesis, the result when a 10\% error is considered.

\begin{table*}
\begin{center}
\caption{Mean errors, mean offsets ($\Delta$) and standard deviation ($\sigma$) of the residuals in dex units when comparing 12+log(O/H) and  12+log(S/H)  as derived from {\sc HCm} with the input values of the {\sc popstar} photoionization models and using the same lines predicted by them. The values between parenthesis correspond to the results when a 10\% error is considered. Different cases attending to to distinct sets of emission lines are listed.}

\begin{tabular}{lcccccc}
\hline
Case   &  \multicolumn{3}{c}{12+log(O/H)} &  \multicolumn{3}{c}{12+log(S/H)}  \\
\hline
  &  Mean error & Mean $\Delta$    &  St.dev. $\Delta$ & Mean error & Mean $\Delta$  &  St.dev. $\Delta$ \\
\hline
All\footnote{Only models which predict a flux for the auroral lines} & 0.000 (0.093) & +0.000 (+0.018) & 0.003 (0.058) & 0.027 (0.040) & -0.011 (+0.007) & 0.095 (0.113)  \\  
Strong lines\footnote{Only models in the adopted O/H-$U$ relation} & 0.000 (0.059) & +0.001 (+0.012) & 0.031 (0.069) & 0.016 (0.021) & -0.003 (+0.018) & 0.138 (0.168)  \\  
Red range$^a$ & 0.002 (0.192) & +0.005 (-0.017) & 0.064 (0.258) & 0.025 (0.039) & -0.003 (+0.005) & 0.099 (0.145)  \\  
Red strong lines$^b$ & 0.000 (0.128) & -0.000 (+0.014) & 0.007 (0.145) & 0.015 (0.024) & -0.013 (+0.020) & 0.132 (0.210)  \\

\label{test}
\end{tabular}
\end{center}
\end{table*}

The resulting values obtained from the comparison between the results from {\sc HCm} with the values in the models shows that we can recover the chemical abundances using only emission-lines with a high precision, although this can be altered when the uncertainty in the measurement of the corresponding lines are also considered. This uncertainty can be due to very different causes, including extinction or aperture factors, but their impact on the results of our method should be considered by the user when include certain emission lines.

\section{Results and discussion}

\begin{figure*}
   \centering
\includegraphics[width=0.45\textwidth,clip=]{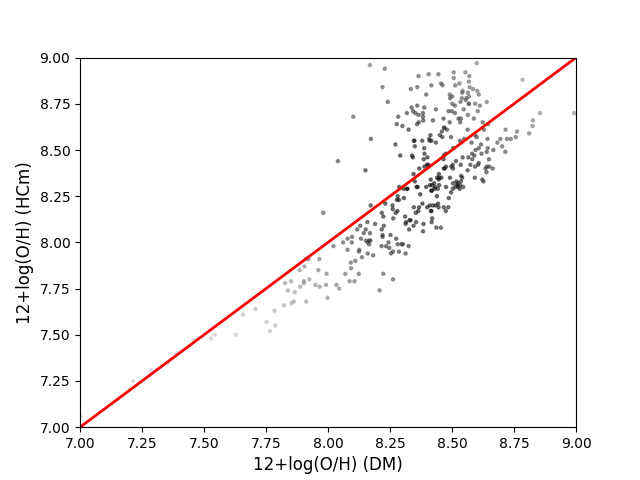}
\includegraphics[width=0.45\textwidth,clip=]{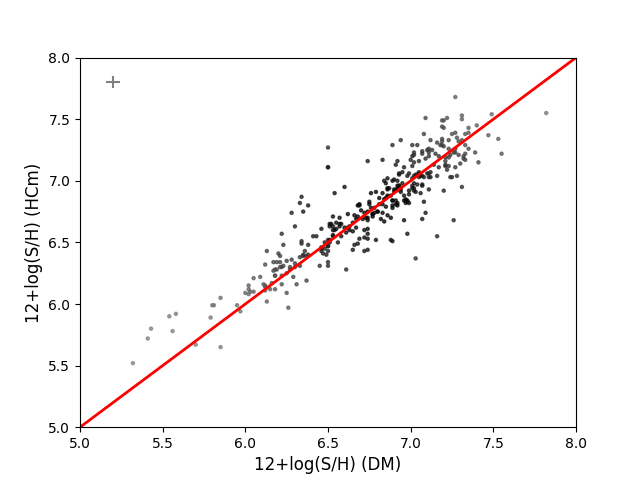}

\caption{Comparison between the chemical abundances derived by HCm
for the selected control sample when at least one auroral line was measured with enough signal-to-noise ratio, At left,12+log(O/H), and at right 12+log(S/H). In both panels $x$ axis represents the total abundances derived following the direct method and $y$ axis the results from {\sc HCm} using all available emission lines as input. The red solid line represents the 1:1 relation.}
\label{comp1}%
\end{figure*}

\subsection{Comparison with the control sample}

In this subsection we describe the results obtained from  our code {\sc HCm}, in its version 6.0, capable of calculating N, O, and S elemental abundances, and we discuss their consistency with the abundances calculated for the same objects using the direct method.
To do so, we employed the sample of objects described in subsection 2.1, which consists of the \hii\ regions in close spiral galaxies and \hii\ galaxies  studied in \cite{dz22}, 
whose total elemental abundances were obtained using at least a measured electron temperature. 

\begin{table*}
\begin{center}
\caption{Mean errors, mean offsets ($\Delta$) and standard deviation ($\sigma$) of the residuals in dex units when comparing 12+log(O/H) and 12+log(S/H) derived from {\sc HCm} and the direct method applied to the whole studied control sample, for disk \hii\ regions (DHR) and \hii\ galaxies (HIIG). Results are given for the case that  all lines, including auroral,  are considered as input, and for the case when  only strong lines are used.}

\begin{tabular}{lccccccccc}
\hline
Type   &  N &  Mean (DM)  &  \multicolumn{3}{c}{HCm (all lines)} &  \multicolumn{3}{c}{ HCm (strong lines)}  \\
\hline
  &   &   & Mean error & Mean $\Delta$    &  St.dev. $\Delta$ & Mean error & Mean $\Delta$    &  St.dev. $\Delta$ \\
\hline
 \multicolumn{9}{c}{12+log(O/H)}  \\
\hline
All & 351  &  8.35 & 0.07  &   -0.03  &  0.19 & 0.05  & +0.06  &  0.29  \\  
DHR & 255  &  8.44 & 0.08  &   +0.01  &  0.23 & 0.05  & +0.03  &  0.29  \\  
HIIG & 96  &  8.10 & 0.05  &   -0.12  &  0.10 & 0.06  & +0.15  &  0.27  \\  
\hline
  \multicolumn{9}{c}{12+log(S/H)}  \\&  \multicolumn{7}{c}{12+log(S/H)} \\
\hline
All & 351  &    6.75  &  0.04  & +0.02  &  0.18 & 0.02  &   +0.06  &  0.29  \\  
DHR & 255  &    6.75  &  0.03  & +0.02  &  0.17 & 0.02  &   +0.10  &  0.29  \\  
HIIG & 96  &    6.75  &  0.06  & +0.05  &  0.21 & 0.02  &   -0.02  &  0.24  \\  
\hline

\label{disp}
\end{tabular}
\end{center}
\end{table*}

In Figure \ref{comp1} we show the comparison between the abundances calculated using the data in \cite{dz22} with the results from {\sc HCm} when all possible lines, including auroral lines, are used as input by the code. We also list in Table \ref{disp} the mean offsets and the standard deviations of these comparisons. As it can be seen, the agreement in both plots is quite good, without any neither  discontinuity nor break over a wide range of metallicity, and only with some little deviations that can be attributed to several factors  including the atomic coefficients, the assumed relations between the electron temperatures associated to each ion, or the assumed ICFs for each element. These factors are arbitrary and can lead to deviations that are usually lower than the observational uncertainties. 

\begin{figure*}
   \centering
\includegraphics[width=0.45\textwidth,clip=]{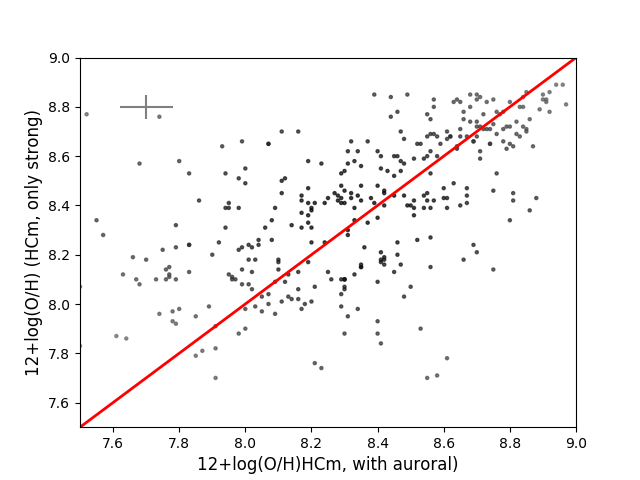}
\includegraphics[width=0.45\textwidth,clip=]{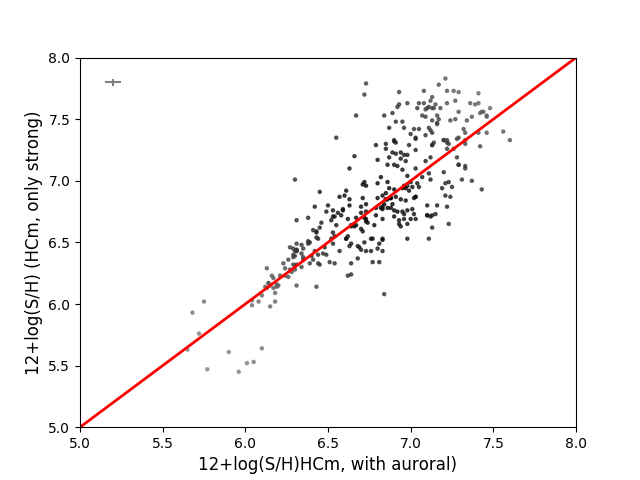}

\caption{Comparison between the chemical abundances derived by HCm
for the selected control sample considering auroral lines (x axis) and without them (y axis), At left,12+log(O/H), and at right 12+log(S/H). The red solid line represents the 1:1 relation.}
\label{comp2}%
\end{figure*}

In absence of any auroral line, which is the most common situation in weak regions or for metal-rich objects, where the cooling is much more efficient and the emissivities of all CELs is lower, the code can also provide a solution with its corresponding uncertainty. However, additional assumptions must be taken in order to provide a solution consistent with the direct method. As discussed in \cite{hcm14}, this is done by assuming an empirical relation between O/H and $U$ to constrain the grids, in such a way that models with lower metallicity have in average higher excitation and, on the contrary, low excitation regions are compatible with metal-rich regions. 
This assumption is strictly empirical and its interpretation, as it will be discussed in next subsection is subject to uncertainties that we partially consider in our code (i.e. by adopting ranges of variation in the adopted empirical relation),}but it helps breaking the degeneracy of certain emission-line ratios such as R23, with metallicity. A similar constrain is considered by the code when N/O cannot be calculated, so [\nii] emission-lines can only be used to derive O/H after assuming a certain arbitrary relation between O/H and N/O. As discussed in \cite{hcm14} and several other works describing {\sc HCm}, this assumption is usually made in all strong-line methods based on {\nii} emission lines in the optical, so it can lead to non-negligible mistakes if the studied object does not follow the expected adopted relation.
In any case, in order to let the user to explore additional uncertainties owing to the adoption of different relations, these can be edited or replaced in the code to account for different families of objects or prescriptions. For instance, as studied in \cite{pm21}, for EELGs, the assumed relation between O/H and $U$ is slightly different, pointing towards slightly higher $U$ values for lower metallicity.  

Table \ref{disp} lists, and Figure \ref{comp2} shows, the results obtained by HCm with and without auroral lines, which implies in the latter case the use of only strong nebular lines. As in the case of other works describing {\sc HCm} we obtain an agreement for most 
of the points in all regimes that it is compatible with the observational errors, although with a larger dispersion than in relation with the case where any auroral line is available. Anyway, it is important to underline the fact that no specific assumption must be taken in relation with the adopted temperature relations or the ICF as these are implicit in the models and the code only searches for total abundances.

A more detailed comment can be given in relation to the values found for 12+log(S/H), not covered in previous versions of the code. We also find for this element very good agreement, even in absence of any auroral line, although with a dispersion around 0.2 dex. This result confirms that our approach to estimate S/H abundances based on the measured [\sii] and [\siii], once O/H and $U$ have been estimated, is basically correct, and that our method is valid to be probed for large samples of objects.  

\begin{figure*}
   \centering
\includegraphics[width=0.4\textwidth,clip=]{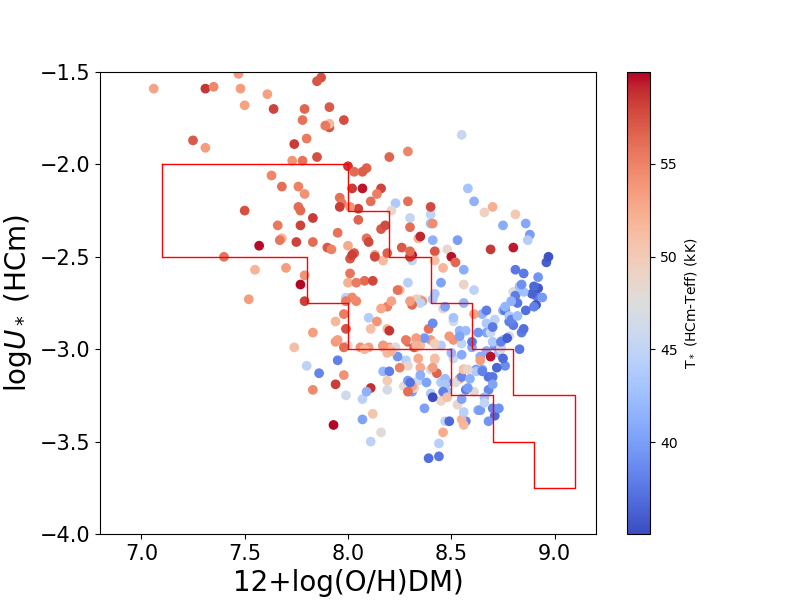}
\includegraphics[width=0.4\textwidth,clip=]{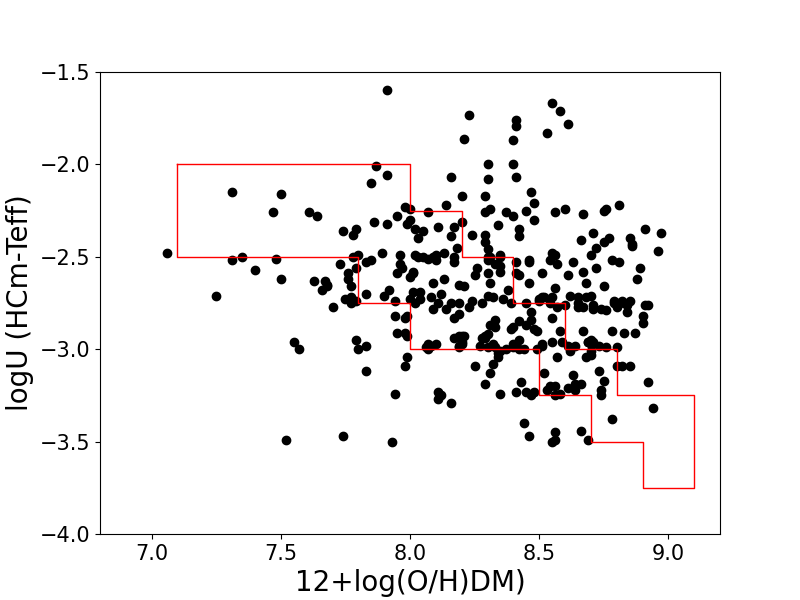}

\caption{Left: Relation between 12+log(O/H) and log $U_*$ ((i.e. the ionization parameter calculated assuming a source of fixed $T_*$, in this case {\sc popstar} model atmospheres) as derived by {\sc HCm} for those objects in the control sample for which at least one auroral line has been used. The solid line encompasses the values for the models used when only strong lines are used as input. The points are colored according to the values for $T_*$ found by {\sc HCm-Teff}. Right: Same relation, but comparing the 12+log(O/H) derived by {\sc HCm} with the log $U$ values found by {\sc HCm-Teff} (using sources of different $T_*$ with {\sc WM-Basic} stellar atmospheres).}
\label{OH-U}
\end{figure*}

\subsection{An interpretation of the empirical relation between O/H and $U$}

One of the foundations of the above described methodology using the {\sc HCm} code to derive chemical abundances when no nebular-to-auroral line ratios  are provided in the input is the assumed empirical anticorrelation between O/H and $U$, what is used by the code constraining the grid of models to avoid the double-valued relations  of certain emission-line ratios with O/H (e.g. R23, \citealt{pagel1979}). Although not universal and highly dependent on the studied class of objects, this assumption comes from a widely observed relation between certain ratios of consecutive emission-line fluxes and the chemical content of the gas.
In this subsection, we try to make a deeper examination on the physical basis of this assumption, taking advantage of the direct relation between certain emission-line ratios with the input parameters considered in our grids of photoionization models.

In left panel of Figure \ref{OH-U} we show the relation between the values of O/H and $U$ found by {\sc HCm} for the objects in the control sample for which
at least one auroral line is used as an input. In this case, no previous assumption is required on the relation between O/H and $U$ to retrieve precise estimates of the chemical abundances, and thus no constrain is made to the used grid of models.

As it can be seen, the code finds for the subsample from \cite{dz22} used in this work 
the expected anticorrelation already found for a larger 
sample in \cite{hcm14}, in such a way that objects with a lower metallicity have in average higher values for log $U$. This relation is represented by the solid line shown in the same figure and also encompasses   The set of models used by the code to perform the calculations when no auroral line is given as input.
In appearance the O/H-$U$ relation derived for our control sample is , somehow less clear than in the case of the sample studied in \cite{hcm14} ($\rho_S$ = -0.48), maybe owing to the higher weight in the sample analyzed here of the objects at large metallicity, for which the trend does not seem to exist, and that  some objects lie out of the limits designed to constrain the grid of models. However, the agreement between the abundances derived by the code when no auroral lines are given using  a constrained grid is quite reasonable, as can be seen in Fig. \ref{comp2} and inspecting the offsets and dispersions listed in Table \ref{disp}.

Focusing on this observed O/H - $U$ anticorrelation itself, there is not a consensus in the literature about if this correlation is real or has a physical basis. On one hand, several authors (e.g. \citealt{de86,maier2006,morisset16,garner25}) have reported the same anticorrelation in samples of \hii\ regions by fitting high-to-low excitation emission-line ratios with photoionization models.
On the other hand, some other authors found a lack of correlation or even a positive correlation between O/H and $U$ (e.g. \citealt{kennicutt1996,dors2011,kreckel2019,zinchenko19}) and/or alert on the strong dependence of the results on the initial assumptions made in the photoionization models used to derive these properties (e.g. \citealt{ji2022}).

In the several papers describing the different versions of {\sc HCm}, it is stated that the derived $U$ values are subject to the definition of the assumed geometry or the adopted input SED, so its final value is just a convention taken to compare heterogeneous samples or to fix an empirical relation between the metal content of the gas and the observed behaviour of emission-line ratios very dependent on the gas excitation, such as O2O3 or S2S3, as explained in the above sections. However, a grid of models exploring a variation of properties such as the shape of the incident SED, the geometry, or the assumed dust-to-gas mass ratio is required to correctly interpret from a physical basis the observed anticorrelation between metallicity and $U$, as all these other properties have a non-negligible impact on the observed emission-line ratios.

It has been demonstrated that making small modifications
 on the assumed incident SED of the models does not imply changes beyond the reported errors for the results from {\sc HCm} on the derived abundances, what is corroborated by the agreement found when these results are compared with those coming from the direct method.
For instance, as explored in \cite{hcm14} variations of the age of the {\sc popstar} \cite{popstar} stellar atmospheres used in the models do not lead to deviations in the derived oxygen or nitrogen abundances larger than 0.02 dex. In addition, in \cite{hcm-agn}, for AGN,  a change in the assumed $\alpha_{OX}$ parameter shaping the incident SED,  do not alter the derived abundances. Finally, in \cite{pm21}, for EELGs, it is proven that  changes in the properties of the assumed BPASS \cite{bpass} model atmospheres, such as  age, IMF or upper mass limit, only imply variations in the derived log $U$, but neither for N/O nor for O/H.

Nevertheless, a closer inspection of the real nature of the derived $U$ values in {\sc HCm} can only be made by means of an alternative approach exploring the nature of the incident field of radiation.
For this reason, in left panel of Figure \ref{OH-U}  we  use the denomination $U_*$ to denote the ionization parameter calculated by {\sc HCm} to underline the fact that the used grid of models do not consider variations of the incident SED (i.e. they are calculated at a around fixed equivalent effective temperature, $T_*$)  so the variation of certain  emission-line ratios usually associated to variations of the excitation (e.g. O2O3, S2S3) are exclusively interpreted by the code as due to changes of $U$ according to this grid of models.

Alternatively, an analysis of $U$ considering variations of the incident field of radiation  can be performed by using the code {\sc HCm-Teff} \cite{hcm-teff}, with a very similar methodology to {\sc HCm}, but using as inputs the metal content of the gas, and at least two emission-line ratios to simultaneously fix $U$  and $T_*$, thus providing an additional perspective to interpret the observed anticorrelation between O/H and the obtained $U_*$ parameter when the SED is fixed. The study of the control sample studied in this work allows this study, as it has been selected to simultaneously have emission-line ratios, O2O3 and S2S3 involved in the softness diagram \cite{pmv09}, what allows {\sc HCm-Teff} to extract both $U$ and $T_*$.

We applied then the {\sc HCm-Teff} code on its version 5.6, assuming in the models incident SEDs of individual WM-Basic massive stars \cite{wmbasic} in the range 30 - 60 kK, with a plane-parallel geometry. More detailed description of the photoionization models and the code are given in \cite{hcm-teff}.
In right panel of Figure \ref{OH-U} we show the relation between the same O/H values represented in left panel, but as compared with the log $U$ values found in this other version of the code that considers possible variations of the hardness of the field of radiation. As it can be seen, the correlation practically disappears. On the contrary, as shown with color of the points in left panel, encoded with the $T_*$ values found by {\sc HCm-Teff} code, it seems that the anticorrelation observed when it is compared with the log $U_*$ values found by {\sc HCm} could be explained in terms of average higher $T_*$ values in low-metallicity objects, and lower $T_*$ sources for the high-$Z$ ones. In fact the Spearman coefficient for the O/H - $T_*$ is -0.65, which is lower than the one obtained for log $U_*$.

The anticorrelation between O/H and $T_*$ was already pointed by \cite{hcm-teff} analyzing the \hii\ regions in NGC~628, M~51, and M~101, where the presence of a negative radial gradient of metallicity were obtained in coincidence with a positive radial gradient of the hardness of the incident field of radiation.
This relation was also indirectly pointed out by other authors (e.g. \citealt{dopita2006,eldridge2011}), who state that the relation observed with $U$ when this is derived using emission-line ratios could be the consequence of a higher opacity  in the atmospheres of the ionizing stars. At same time, other empirically found relations between metallicity and other properties of the ionizing stellar population have been found in other works, such as Initial Mass Function (IMF, \citealt{martinnavarro2015}), SFR or sSFR (e.g. \citealt{dopita2014, mingozzi2020}), or age \cite{pellegrini2020}, underlining the anticorrelation between metallicity and the hardness of the incident field of radiation, and not for the ionization parameter.

Nevertheless, a more precise and robust characterization of the relation between the metal content of the ionized gas with the ratios of fluxes emitted by different ions, should also cover all physical mechanisms other than $U$ and $T_*$, including dust-to-gas mass ratio or gas geometry. In any case, the functional observed anticorrelation between O/H and log $U_*$, that implicitly includes all these factors as they are not varied in the models, is still useful, given that, firstly,  it allows a precise determination of chemical abundances, reducing the number of explored models in the grid, and, secondly, it also allows a consistent comparison between objects in conditions where not all the mechanisms affecting the excitation  can be independently fixed.

\begin{figure}
   \centering
\includegraphics[width=0.45\textwidth,clip=]{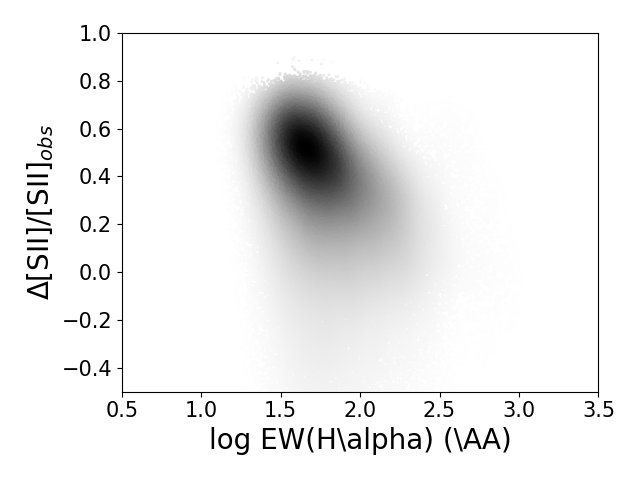}

\caption{Density scatter plot representing the relation between the derived [\sii] flux excess relative to the observed emission as a function of EW(\ha) for the selected sample of pointings in MaNGA.}
\label{dig}%
\end{figure}

\begin{figure}
   \centering
\includegraphics[width=0.45\textwidth,clip=]{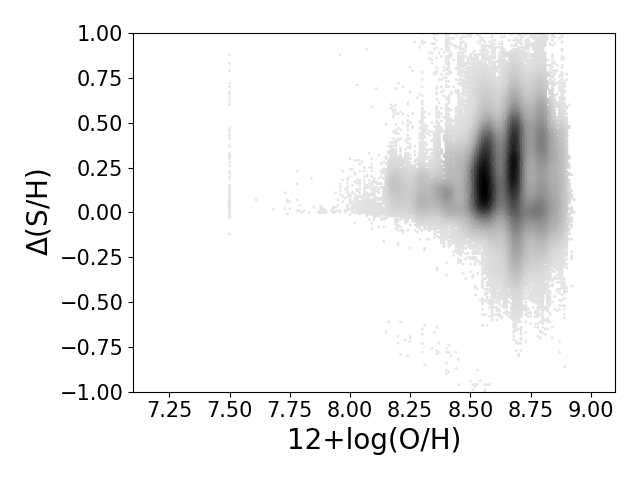}

\caption{Density scatter plot for the selected sample of pointings in MaNGA representing the difference between the total sulphur abundance as derived by {\sc HCm} using the measured [\sii] lines, minus the same abundance calculated using the [\sii] lines corrected as described in the text. This is plotted as a function of 12+log(O/H) as derived using the same code.}
\label{Diff-S}%
\end{figure}

\subsection{Correcting the [\sii] emission excess in MaNGA}

One of the drawbacks that the S abundance derivation may have in samples of objects with very low spatial resolution comes from the confirmed fact that a non-negligible fraction of the collected [\sii] emission presents an excess in relation to what it would be expected just accounting for the gas ionized by a resolved young massive stellar clusters (e.g. \citealt{zurita2000,haffner09}).
This diffuse ionized gas (DIG) emission  may have  a non-negligible impact on the determination of chemical abundances (e.g. \citealt{kumari19}),
having a non-negligible weight  in different galaxy surveys based on integral field spectroscopy like CALIFA \citep{lacerda18} or MaNGA \citep{belfiore16}. 
Moreover, the impact of the excess in the [\sii] emission on the determination of abundances was already documented in  the specific case of the sample of selected MaNGA pointings studied here by \cite{z21}, who realized that the N/O abundance ratios derived using {\sc HCm} were much lower when the code was fed with [\sii] emission lines, allowing that both  N2O2 and N2S2 parameters were simultaneously used.
Since the code uses these two emission-line ratios as source of information to provide a solution for N/O \citep{pmc09}, this implies that the [\sii] emission is overestimated in relation to what would correspond to the N/O abundance as solely traced by N2O2, relatively unaffected by DIG emission \citep{zhang17}.

Several recipes have been proposed to correct the emission from DIG in galaxy disks with star formation (e.g. \citealt{valeasari19}), but a consistent criterion that might be applied to a large number of objects like those in MaNGA without a detailed study of the geometry or the source of the incident field of radiation in each region is not straightforward \citep{weber19}. 
As an alternative, even though  DIG emission also contributes  to the emission of other low-excitation lines such as [\nii] (e.g. \citealt{blanc09}) or H recombination lines, we can  consider that the excess due to the DIG in these lines is much lower than in the case of [\sii] in our sample of star-forming dominated selected objects. 
This is partially confirmed for the sample studied here by the results from \cite{pm23}, who studied the MaNGA sample in the context of the softness diagram and showed that the hardness of the field of radiation derived for these selected \hii\ regions is higher than the maximum $T_*$ considered for massive stars in a 60\% of the sample when the S2S3 ratio is used. Instead, this proportion is lowered to values compatible with other \hii\ regions observed using narrow long-slit spectroscopy, such as CHAOS \cite{chaos-n628}, when [\nii] is used replacing [\sii].

Therefore, in order to provide a more precise derivation of the S abundance in the MaNGA sample, it is fundamental to quantify and correct the [\sii] excess, as [\sii] lines are necessary as input for our code to derive S, while this line is not used to derive other abundances, like O/H or N/O, whose results are then much less affected by this contribution to other lines.
In this way, assuming that the lines involved in the N2O2 parameter are  much less affected by DIG emission, we can minimize and correct the DIG emission of [\sii], 
by taking the N/O values derived by {\sc HCm}  when only [\nii] and [\oii] are used. Later,  these abundance ratios can be utilized to derive the fraction of [\sii] flux not affected by DIG by means of the relation of them with the N2S2 parameter.
 as predicted by the photoionization models presented in \cite{hcm14}. The functional form of this fit is:

\begin{equation}
log(N/O) = 0.9924 \cdot N2S2 - 0.8484 
\end{equation} 

\noindent where N2S2 = log([\nii] $\lambda$ 6584 \AA/[\sii] $\lambda$ 6717+31 \AA). This relation has a Pearson statistical significance r = 0.992 and a p-value = 0.0.
When models considering a S/O ratio different than the solar value are also examined, this function does not lead to [\sii] fluxes deviating more than a 3\% in relation with the fluxes derived by this fit, what is much less than the measured excess
in most of the sample. We then took for our analysis the resulting corrected [\sii] derived from this relation.

Several facts on the resulting corrected [\sii] fluxes point to the DIG nature of the derived excess. Firstly, most of the   analyzed MaNGA objects (82\%) show a measured [\sii] emission that is larger than their corresponding  corrected derived flux under the hypothesis of no DIG contribution based on the measured [\nii] emission. For the rest of the objects for which the corrected value is larger than the measured one, the average difference is less than a 5\%.
Secondly, as it can be seen in Figure \ref{dig},  the [\sii] excess presents a negative correlation with the \ha\ equivalent width, that it is known to be lower in those regions with a larger DIG emission (e.g. \cite{lacerda18}).
Though our sample does not contain pointings dominated by DIG emission, as EW(\ha) is larger than 3 \AA\ in all cases,
the relative contribution of the excess is in average larger than a 20\% in those pointings with EW(\ha) $<$ 100 \AA, while for the most star-forming dominated points, the [\sii] correction is quite similar to the measured value.

We can then use the new [\sii] values to calculate S abundances replacing the measured ones. Although this does not ensure that all DIG contamination is removed from our measurements, it substantially reduces the contribution to the same level as in other samples observed with higher spatial resolution, such as in {\sc CHAOS}, as explained in \cite{pm23}.
For our final calculations we took the corrected [\sii] flux, as derived using the above recipe based on the previous calculation of the N/O abundance ratio, only in those pointings where this corrected value is lower than the measured one. For the rest, we still used this latter, which are mainly those pointings with a very high value of EW(\ha).

In order to illustrate the impact of this correction on S abundances, we show in Figure \ref{Diff-S}, the difference between the total S abundances  using the measured and the corrected [\sii] lines calculated from {\sc HCm}. The difference is presented  as a function of the total oxygen abundance, even for those objects for which the derived S/H abundance based on the  corrected [\sii] line is larger than the value based on the measured one.
As it can be seen, there is a slight correlation to find larger deviations for higher metallicities, and in the over-solar regime the S abundances can be in average 0.3 dex larger  if the [\sii] excess is considered, underlying the importance of this correction even for star-forming dominated objects.

Interestingly, considering that in most of the galaxies of this sample there is a negative radial gradient of metallicity \citep{z21}, this could be contradictory to the studies stating that the DIG emission is larger in the outer galactic parts (e.g. \citealt{belfiore17}). However, as we only consider star-forming dominated pointings with very high EW(\ha) values, this can be only interpreted in terms of more concentrated evolved stellar populations in the galactic central parts, where stellar density is also higher \citep{morisset16,erroz19}.

\begin{figure*}
   \centering
\includegraphics[width=0.7\textwidth,clip=]{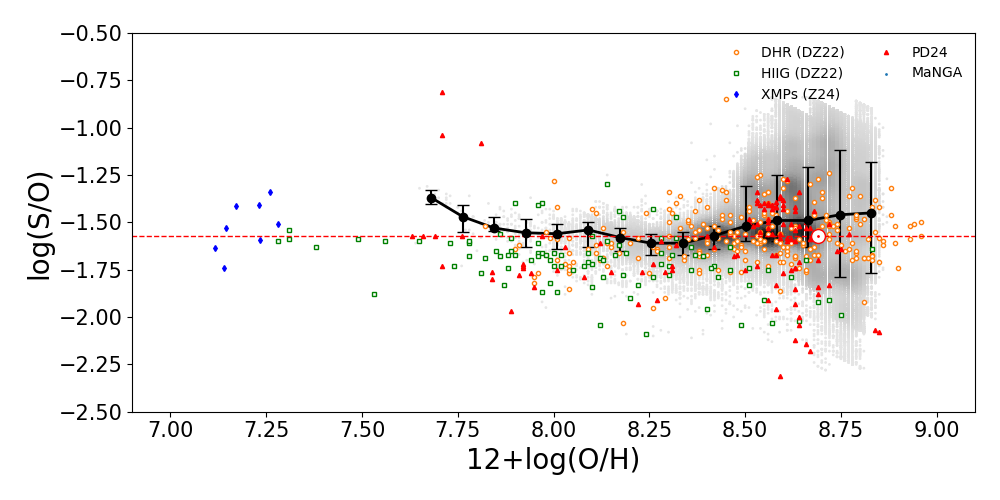}

\caption{relation between 12+log(O/H) and log(S/O) as derived by {\sc HCm} for all the samples analyzed in this paper, including the {\sc MaNGA} selected star-forming pointings (grey density points with median values represented as black circles joined by solid black line), disk \hii\ regions (orange circles) and \hii\ galaxies (green squares) from \cite{dz22}, XMPs (blue diamonds) from \cite{zinchenko24}, and the star-forming regions analyzed in the IR (red triangles, \citealt{pd24}). The horizontal line and the solar symbol represent the solar values given by \cite{asplund09}.}
\label{OH-SO}%
\end{figure*}

\begin{figure*}
   \centering
\includegraphics[width=0.7\textwidth,clip=]{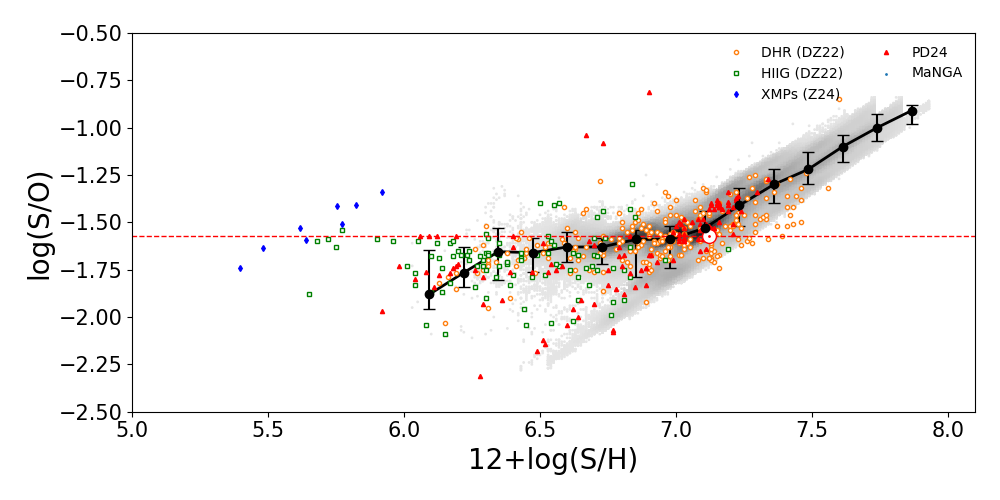}

\caption{Same Figure as Figure \ref{OH-SO}, but representing log(S/O) as a function of 12+log(S/H).}
\label{SH-SO}%
\end{figure*}

\subsection{The relation between metallicity and S/O}

Once we have a methodology to derive sulphur abundances that can be applied to a large and representative sample of objects, regardless of the fact that the direct method can be used, we examine the overall behaviour of the relation between metallicity and S/O ratio over a wide range of metallicity. Even though the MaNGA sample presents certain objects for which both [\oiii] and [\siii] auroral lines could be measured with enough signal-to-noise ratio (around 1131, a 0.56\% of the sample), we only considered strong nebular lines for the whole sample for our calculations, under the above discussed assumption of an empirical relation between O/H and log $U_*$. Anyway, we checked that the derived abundances in this sample using the auroral lines did not deviate more than the values already discussed. 

Figure~\ref{OH-SO} displays the relation between the sulphur-to-oxygen abundance ratio, log(S/O), and the oxygen abundance, 12 + log(O/H), as derived by \textsc{HCm} for the MaNGA sample of star-forming selected pointings. Each blue dot represents a pointing in which all required emission lines were measured with sufficient signal-to-noise. A grey density map highlights the statistical distribution of the points. while the black circles joined by a black line represents median values per bin in the sample. A red dashed horizontal line marks the solar S/O value (log(S/O)~$= -1.57$, \citealt{asplund09}), and a red dot indicates the solar oxygen abundance (12 + log(O/H)~$= 8.69$). 

The figure also includes comparison samples from \cite{dz22}, based on the direct method: classical disk \hii\ regions (DHR, orange circles) and high-ionization \hii\ galaxies (HIIG, green squares). We also include  the sample of star-forming objects studied by \cite{pd24} using {\sc HCm-IR} and mid-IR emission-lines (red triangles).
In addition, since  the low-metallicity regime (12+log(O/H) $<$ 7.6) is poorly populated, 
by the MaNGA sample and also very few \hii\ galaxies taken from \cite{dz22},
we also included the sample of Extremely Metal Poor Galaxies (XMPs; O/H $<$ O/H$_{\odot}$/10) studied by \cite{zinchenko24} (blue diamonds) to cover this range.

In this regime, we observe some scatter that, considering the uncertainties in the estimations, is compatible with the S/O solar ratio, 
 although a number of individual points show enhanced log(S/O), a trend which is also traced in the {\sc MaNGA} sample. These outliers may reflect local chemical enrichment by Type II supernovae or other stochastic effects in metal-poor galaxies. Alternative scenarios include IMF variations or contributions from very massive stars \citep[e.g.,][]{goswami24}.
In the range 7.6 $<$ 12+log
(O/H) $<$ 8.0, both samples from {\sc MaNGA} and \cite{pd24}, present a non-negligible number of objects with S/O ratios slightly over solar. However, when considering all data, the majority of the objects show an almost solar S/O ratio. 
This behavior is consistent with the expectation that both oxygen and sulphur, being produced mainly by core-collapse supernovae, evolve in lockstep.

At very high metallicity (12+log(O/H) $>$ 8.5),  the scatter increases  and, some {\sc MaNGA} regions present sub-solar S/O ratios, mimicking the trend observed in certain objects in  \citep{pd24}."
However, there is also a considerable amount of {\sc MaNGA} regions that show a S/O ratio above the solar value, coinciding with the trend reported by \cite{dz22}. 
This increase could be partially due to progressive oxygen depletion onto dust grains, given that sulphur is less refractory than oxygen \citep[e.g.,][]{garnett89, henry93}.

As suggested by \cite{dz22}, we can minimize the dependence of total metallicity on its depletion onto dust, by representing its relation with S/O using total sulphur abundance, which is thought to not to be affected by this. This is shown in Figure \ref{SH-SO} for the same samples described above, along with the corresponding S/O value using again a red dashed line and the solar S/H value at 7.12 \citep{asplund09}.

We note that this plot is consistent with the previous one, as the sample from \cite{zinchenko24} is clearly differentiated from the others by covering the low-metallicity regime in S/H. 
Interestingly, we observe to major trends. In the range 5.9 $<$ 12+log(S/H) $<$ 6.5, there is essentially no trend, observing that objects cluster around slightly sub-solar S/O values, and without any object with oversolar S/O values in this regime, what could point to that the objects simultaneously showing low O/H and high h S/O ratios could have very large dust contents. On the other hand, although DHRs and IR data tend to cluster around the solar S/O ratio for 12+log(S/H) below 7.12,  there is a clear trend to find  increasing log(S/O) for increasing 12+log(S/H) at higher abundances. In addition, the {\sc MaNGA} sample seems to present several behaviors.  For over-solar S abundances, the same increasing trend is observed, while  for sub-solar S abundances, there are two trends: one that, as in the previously described samples, has average values clustering around the solar S/O ratio, and another one showing the same increasing S/O values  observed in the high-metallicity range.

Thus, our results  indicating that the increase of log(S/O) with the total sulphur abundance is much clearer than when it is studied as a function of total oxygen abundance, is an evidence of the growing  depletion of this element into dust grains for higher metallicity, underlining that total sulphur abundance can be a better tracer of the overall metal content in these objects, and giving important clues on the amount of oxygen retired from the gas-phase. 
This interpretation is reinforced by the fact that the {\sc MaNGA} data has been corrected for the [\sii] excess, which is particularly relevant at high metallicities. As shown in Figure~\ref{Diff-S}, and discussed by \citet{pm23}, DIG can lead to overestimated [\sii] fluxes in IFS data, and thus to lead to artificially higher sulphur abundances. The application of a correction based on the model-based N2S2--N2O2 relation mitigates this effect, reducing potential overestimates in log(S/O) by up to 0.3~dex at solar and over-solar O/H. Without such correction, the observed increase in log(S/O) at high metallicity would likely be stronger, overestimating the depletion issue for oxygen.

In any case, the fact that the {\sc MaNGA} distribution shows overall a broad agreement in the global trend with the results from the direct-method or IR samples, likely reflects the model-based and internally consistent nature of \textsc{HCm}, which avoids external assumptions on ICFs or temperature relations, what reinforces the reliability of our method.

\section{Summary and conclusions}

We presented a new application of the \textsc{HCm} code in its version 6.0 to derive sulphur abundances in ionized gas regions using optical emission lines, with particular focus on the behaviour of the S/O ratio with metallicity. This version of the code incorporates updated photoionization model grids with variable S/O, and a robust treatment of strong-line diagnostics that allows consistent derivation of O, S, and N abundances even in the absence of auroral lines.

Our main conclusions are as follows:

\begin{itemize}
  \item The new version of \textsc{HCm} provides S/H abundances that are consistent with those derived using the direct method, with typical offsets below 0.05~dex and dispersions compatible with the observational uncertainties. As in previous versions, a good agreement between the abundances derived using the direct method and those calculated by {\sc HCm} can be reached even in absence of any auroral line assuming an empirical relation between metallicity and the ionization parameter.
   \item We demonstrated that the relation between O/H and U has a functional character that helps to derive abundances, but its physical meaning is much more related with a relation between metallicity and the hardness of the incident field of radiation, as already pointed out by \cite{hcm-teff} using {\sc HCm-Teff}. However, the assumption by {\sc HCm} of an empirical anti-correlation between O/H and $U_*$ (i.e. the ionization parameter calculated with a fixed $T_*$ source) results useful to find out abundances without assuming multiple ionization sources in the models. 
  \item For the {\sc MaNGA} sample, we applied DIG corrections to the [\sii] emission using a model-based relation between N2S2 and N2O2. This step proved crucial to avoid artificial enhancements of S/O at high metallicity.
  \item We find for both {\sc MaNGA} and other control samples that the S/O ratio remains approximately constant and near the solar value across a broad metallicity range, particularly for 8.3~$<$~12+log(O/H)~$<$~8.7. A mild upward trend is observed at over-solar metallicities, much clearer when the relation is represented as a function of 12+log(S/H), possibly due to oxygen depletion onto dust grains.
  \item At low metallicities (12+log(O/H)~$<$~8.0), we observe greater scatter and a population of objects with elevated log(S/O), consistent with previous findings from IR-based analyses. These outliers may be linked to stochastic enrichment processes or variations in stellar yields. However, the absence of these same points when S/O is compared with S/H, an indicator of total metallicity which is not affected by dust depletion,
could indeed point out towards a large depletion factor in these objects. 
\end{itemize}

This study demonstrates that the new implementation of \textsc{HCm} can be reliably used to study sulphur abundances in large IFS surveys, and provides a solid framework to explore the physical processes that modulate the relative enrichment of $\alpha$-elements in galaxies.

\begin{acknowledgments}
This work has been funded by project Estallidos8 PID2022-136598NB-C32 and PID2022-136598NB-C33
(Spanish Ministerio de Ciencia e Innovacion).
We also acknowledge financial support from the Severo Ochoa grant CEX2021-001131-S funded by MCIN/AEI/ 10.13039/501100011033.
EPM also acknowledges the assistance from his guide dog Rocko without whose daily help this work would have been much more difficult.
IAZ acknowledges funding from the Deutsche Forschungsgemeinschaft (DFG; German Research Foundation)---project-ID 550945879
\end{acknowledgments}


\bibliographystyle{aa}
\typeout{}
\bibliography{HCm-opt-sulphur_OJAp}


\end{document}